# Sterile neutrinos: the necessity for a 5 sigma definitive clarification.

Carlo Rubbia[1,2], Alberto Guglielmi[3], Francesco Pietropaolo[3], Paola Sala[4]

[1]GSSI, L'Aquila, Italy,
[2]CERN, Geneva, Switzerland,
[3]INFN, Sezione di Padova, Italy,
[4]INFN, Sezione di Milano, Italy.

*Abstract.*

Several different experiments have hinted to the existence of "anomalies" in the neutrino sector, implying the possible presence of additional sterile neutrinos or of other options. A definitive experimental search, capable to clarify either in favour or against all these anomalies at the appropriate $> 5\sigma$ level has been proposed by the ICARUS-NESSIE Collaboration. The technique is based on two innovative concepts, namely (1) a large mass Liquid Argon Time Projection Chamber (LAr-TPC) now in full operation at LNGS and (2) the search for spectral differences in two identical detectors at different distances along the (anti-) neutrino line(s).

The recent result at CERN-LHC with the detection of the Higgs particle [1] has demonstrated once again the necessity of a more than "5 sigma" evidence for the definitive assessments of fundamental physics discoveries.

As well known, neutrino oscillations have established a picture consistent with the mixing of three physical neutrino $\nu_e$, $\nu_\mu$ and $\nu_\tau$ with mass eigenstates $\nu_1$, $\nu_2$ and $\nu_3$. In particular the mass differences turn out to be relatively small $\Delta m^2_{31} \approx 2.4 \times 10^{-3}$ eV$^2$ and $\Delta m^2_{21} \approx 8 \times 10^{-5}$ eV$^2$. Note that these are differences and that masses are unknown. The sum of the strengths of the coupling of different $\nu$ is very close to 3 [2]. But only assuming that neutrinos in similarity to charged leptons, have unitary strengths that we conclude that the *resulting number of neutrino kinds is 3*. The experimentally measured weak coupling strengths are only poorly known, leaving lots of room for alternatives.

Since the initial remark of Pontecorvo [3] the question of the existence of additional "sterile" neutrinos with masses differences largely in excess than the ones of the standard description has been amply investigated experimentally. Cosmological analyses, and in particular the recently presented results from the PLANCK Collaboration [4] have attracted further attention. These are *"anomalies"* at the few standard deviations level, which, if further confirmed experimentally, could be due to the presence of additional, larger mass differences squared with mixings or of other effects. However the instrumental origin of some or even of all effects cannot be completely excluded at the present stage. The most popular direction is presently the one of "sterile neutrinos" although also other alternatives are possible. These experiments may all point out to the possible existence of at least one fourth non standard and heavier neutrino state driving oscillations at a small distances, with of the order of $\Delta m^2_{new} \approx 0.1\text{-}10$ eV$^2$ and relatively large mixing angle of order $\sin^2(2\theta_{new}) \approx 0.01\text{-}0.1$. Neutrino may then have significant contributions to the Dark Mass.

Two distinct classes of anomalies have been reported, namely:
(1) Observation of a presumed excess signals of electrons from neutrinos from particle accelerators. A first positive 3.8 $\sigma$ signal from the LSND [5] $\bar{\nu}_\mu \rightarrow \bar{\nu}_e$ experiment at an average distance $L/E_\nu \approx$ 0.4-1.0 m/MeV has been only partially confirmed by the MiniBooNE experiment [6] with a positive signal above background but at larger $L/E_\nu \approx$ 1-3 m/MeV corresponding to $E_\nu <$ 500 MeV, with combined evidence of 3.8 σ for the sum of $\nu_\mu$ and $\bar{\nu}_\mu$ signals, while the signal with $E_\nu >$ 500 MeV is not confirmed. ICARUS [7] and OPERA [8] however do not seem to confirm such low energy MiniBooNE excess, implying that for instance the future MicroBooNe result may be instrumental. Therefore there is a substantial tension amongst the findings, at least for the "sterile" assumption. It should be incidentally pointed out that the LSND is very different than others experiments; for instance the LSND effect may be due to a very hypothetical CPT violation at the level of $\approx 2 \times 10^{-3}$ coming from the large $\nu_e$ component produced by the decaying $\mu^+$, oscillating as $\nu_e \Leftrightarrow \bar{\nu}_e \rightarrow e^+ + n$, followed by the neutron capture signal.

(2) An apparent disappearance signal in the $\bar{\nu}_e$ events detected from near-by nuclear reactors [9] and from Mega-Curie k-capture calibration sources [9] in the Gallium experiments to detect solar $\nu_e$. The ratio R between the observed and predicted $\bar{\nu}_e$ reactor signals from previously published results has been recently re-evaluated for all reactor antineutrino spectra with an increased flux. With such a new evaluation, R has decreased to R = (0.927 ± 0.023), leading to a deviation of 3.0 σ from unity. The SAGE and GALLEX experiments [10] record a $\nu_e$ signal produced by k-capture of $^{51}$Cr and $^{37}$Ar, giving R = (0.86 ± 0.05), or about 2.7σ from R = 1.

A recent cosmologic search for neutrino-like *relativistic* particles $N_{eff}$ has been reported by the PLANCK collaboration [4], $N_{eff} = 3.36 \pm^{+0.68}_{-0.64}$, (95%; PLANCK +WP+highL) which however may be increased to $N_{eff} = 3.52 \pm^{+0.48}_{-0.45}$ (95%; PLANCK +WP+highL+H$_o$+BAO). This demonstrates the extreme complexity of the situation.

New and definitive experimental searches, capable to clarify either in favor or against all these anomalies at the appropriate > 5 σ level, are therefore highly desirable. Such an experiment has been proposed by the ICARUS-NESSIE Collaboration at CERN [11] and it is based on two main, innovative concepts and a low energy (E ≈ 3 GeV) neutrino and antineutrino beam from the SPS-CERN Accelerator.

The first new concept is the comparison for spectral differences of the signatures of two (or more) identical neutrino and anti-neutrino detectors located at two different neutrino decay distances. In absence of oscillations, apart some beam related small spatial corrections, the two event distributions will be a precise copy of each other, independently of the specific experimental event signatures and without any Monte Carlo comparison. Therefore an observed proportionality between the two spectra in the different channels proves the absence of additional neutrino oscillations over the measured interval of $L/E_\nu$.

The second new concept is the advent of the novel, now fully operational large mass Liquid Argon Time Projection Chamber (LAr-TPC) detector of the ICARUS collaboration. In a transverse drift field of 500 V/cm with a drift length of 1.5 m the free electrons of ultrapure LAr (≤ 60 parts per trillion of O$_2$ equivalent, free electron lifetime > 5 ms) induce non destructive induction signals in several, subsequent, transparent wires arrays with 3 mm pitch oriented in the direction of the required view. The electron charge is at the end accurately recorded by the collection wire plane. The signal to background ratio for a minimum ionizing track is about 10:1. The electron signal reduction due to the recombination and for the longest drift length is about 17%. The detector is a sampling, homogeneous calorimeter with excellent accuracies and the total energy reconstruction of the event from charge integration. The momentum of escaping high energy muons is well measured by multiple scattering with a resolution of 18% in the required range between 0.8 and 4 GeV/c and a track length > 2.5 m.

The ICARUS experiment has been smoothly operating for about 3 years in the LNGS underground laboratory at 730 km with the CERN neutrino beam [12]. It has collected several thousand neutrino interactions presently under analysis. The large mass detector can be transported to CERN after decommissioning at LNGS [11], ensuring a continued operation in the new experiment. The present ICARUS detector will be complemented with another LAr-TPC detector of ¼ mass and a closer distance from the neutrino beam.

This experiment introduces important new features, which should allow a definitive clarification of all the above described "anomalies":
- $L/E_v$ oscillation path lengths to ensure appropriate matching to the $\Delta m^2$ window for the expected anomalies;
- imaging detectors capable to identify unambiguously all reaction channels with a "Gargamelle bubble chamber class", the LAr-TPC [12];
- magnetic spectrometers to determine muon charge and momentum [13];
- interchangeable ν and anti-ν focussed beams;
- very high rates due to large masses, in order to record relevant effects at the % level (>$10^6$ νμ, >$10^4$ νe, for one nominal year);
- both initial νe and νμ components well identified.

This experiment will collect a large amount of data both with neutrino and antineutrino focussing and the muon momentum determination both inside LAr-TPC by multiple scattering and outside with an iron magnetic spectrometer. The expected CC-event rates *in the absence of "anomalies"* for the ICARUS detector at 1.6 km are shown in Figure 1. The energy distribution of the observed events is extremely similar to the one at the shorter distance (d = 460 m), except however ≈ 20 times the event rate due to its closer position. With one year of neutrino and two years of antineutrino running at the nominal proton energy of 100 GeV and with about 4.5 x $10^{19}$ protons on target [11] we should give a definitive answer to the four following clarifying queries:
(1) LSND/MiniBooNe reactions both for the $v_\mu \rightarrow v_e$ and $\bar{v}_\mu \rightarrow \bar{v}_e$ channels are separately identified with total statistics of > $10^4$ recorded events/y over the $0.4 < L/E_v < 3.2$ m/MeV window. The present method, unlike in the case of LNSD and MiniBooNE, determines separately both the mass difference and the value of the mixing angle. They are detectable with very different and clearly distinguishable patterns depending on the actual ($\Delta m^2 - \sin^2 2\theta$) values. In Figure 2 we show the event rates adding the background to with and without oscillations, $E_v < 5$ GeV, 4.5 $10^{19}$ pot (1 y), d=460 m and d=1600 m and the optimal LSND effect "predictions" $\Delta m^2 = 0.4$ eV$^2$, $\sin^2(2\theta) = 0.01$ [11]. A sterile $v_\mu \rightarrow v_e$ oscillation signal of ≈1200 events should be added above ≈ 5000 background events. The expected excess of events for several other different values of ($\Delta m^2$, $\sin^2(2\theta)$) plane is shown in Figure 3 and summarized in Table 1.
(2) The Mega-Curie k-capture and Reactor induced claimed oscillatory disappearance signals, both for $v_e$ and $\bar{v}_e$ events. The distance of 1600 m and an energy of 3 GeV are equivalent to a $L/E_v$ corresponding for instance to 1.6 m at 3 MeV, very hard limit for reactor experiments. However the energy and the distances of the claimed oscillatory signal the present experiment are optimal for the neutrino beam (see the analogue to Figure 4 and the $L/E_\nu$ with the νe signal).

(3) The νμ event rate in the far position is shown in Figure 1 and it is of about two orders of magnitude greater than νe signal. An oscillatory disappearance analogous to the one of point (2) but so far very poorly known [14] may also be observable in the νμ signal, This is an early experiment because of the large statistics, the analog of the νe disappearance with Reactors and k-capture experiments. The expected distribution is shown in Figure 3 for different values of $\Delta m^2$ and $\sin^2(2\theta) = 0.15$ in order to be near to the value R = 0.927± 0.023 from reactor experiments and the $\bar{\nu}_e$ channel [9]. The corresponding estimated accuracy of the result, including a ≈ 1 % systematic correction, is well in excess of 10 σ in a widely expected $\Delta m^2$ interval.

(4) Accurate comparison between neutrino and antineutrino related oscillatory anomalies with the help of the observation of the final sign of the charge of the muon (CPT violation ?).

In conclusion these very many issues under consideration may be explored in *one single experiment* with the unambiguous identification of the relevant channels (appearance and disappearance simultaneously, neutrino and antineutrino, all final channels). It should then be possible to clarify all these anomalies in a definitive way at the required high level of the statistical and systematic precision in a few years of data-taking.

### Acknowledgements.

The contributions of the ICARUS, NESSIE and CERN-CENF groups and in particular of Sandro Centro, Daniele Gibin, Marzio Nessi, Claudio Montanari, Jan Kiesel and Luca Stanco are warmly acknowledged. The development of the LAr-TPC technology with the ICARUS programme and its deployment in the neutrino beam of an ultrapure Argon mass in excess of 700 tons could not have been realized without the constant support and encouragement of the INFN.

**Table.**

| *Data set #* | 4 | 5 | 6 | 7 | 8 | 9 |
|---|---|---|---|---|---|---|
| $\Delta m^2$ (eV$^2$) | 0.21 | 0.37 | 0.5 | 0.8 | 1.5 | 3.5 |
| sin2(2θ) | 0.03 | 0.01 | 0.005 | 0.0033 | 0.002 | 0.0025 |
| νe total at 460 m | 18614 | 18618 | 18545 | 18853 | 19403 | 22278 |
| νe excess at 460 m | 637 | 641 | 568 | 876 | 1426 | 4301 |
| νe total at 1600 m | 8340 | 8054 | 7668 | 7881 | 7688 | 7492 |
| νe excess at 1600 m | 2031 | 1745 | 1359 | 1572 | 1379 | 1183 |
| **σ of effect (no systematic)** | **22.2** | **19.4** | **15.5** | **17.7** | **15.7** | **13.7** |

Total number of νe events, in the absence of the LNSD like effect, are as follows: 17977 at 460 m and 6309 at 1600 m. Data are for one nominal year of neutrino running and 4.5 x 10$^{19}$ protons on target at 100 GeV.

# Figures.

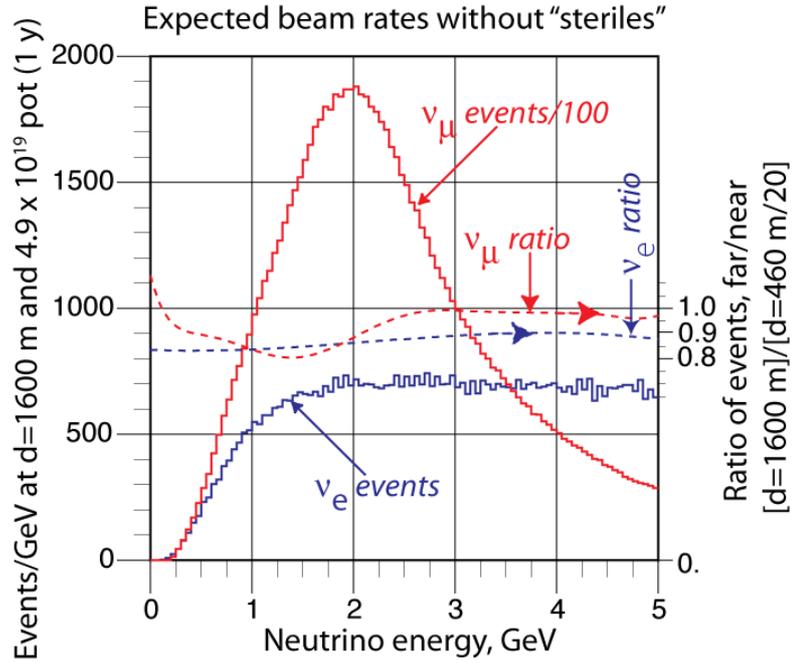

**Figure 1.** Expected neutrino events rate (left graph) without "anomalies" and the ratio of events (right graph) between the far (d = 1.6 km) and the near (d =460 m) positions. The ratio of events shows an extremely weak energy dependence.

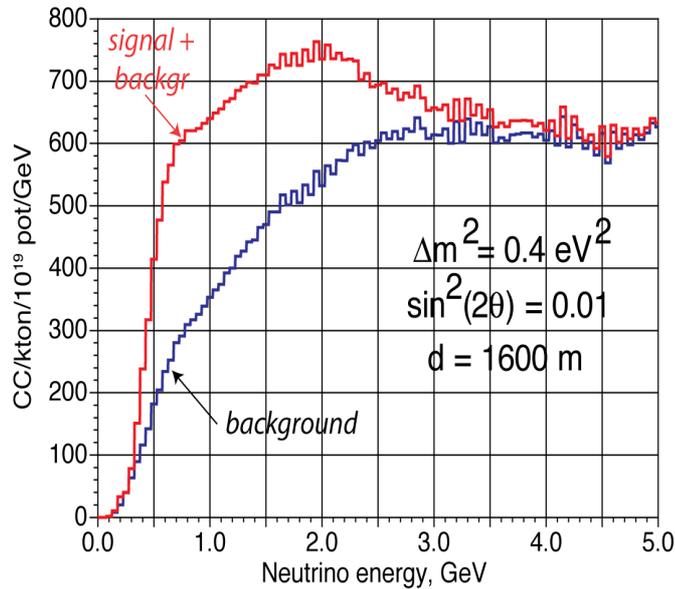

**Figure 2**. [Background + oscillations] and [background] $\nu_e$ events at d = 1600 m and the prediction, $\Delta m^2 = 0.4$ eV$^2$, $\sin^2(2\theta) = 0.01$ [11]. A sterile $\nu_\mu \to \nu_e$ oscillation signal of ≈1200 events is expected above 5000 background events in the position d = 1.6 km, E < 5 GeV and $4.5 \times 10^{19}$ p.o.t. (1 y).

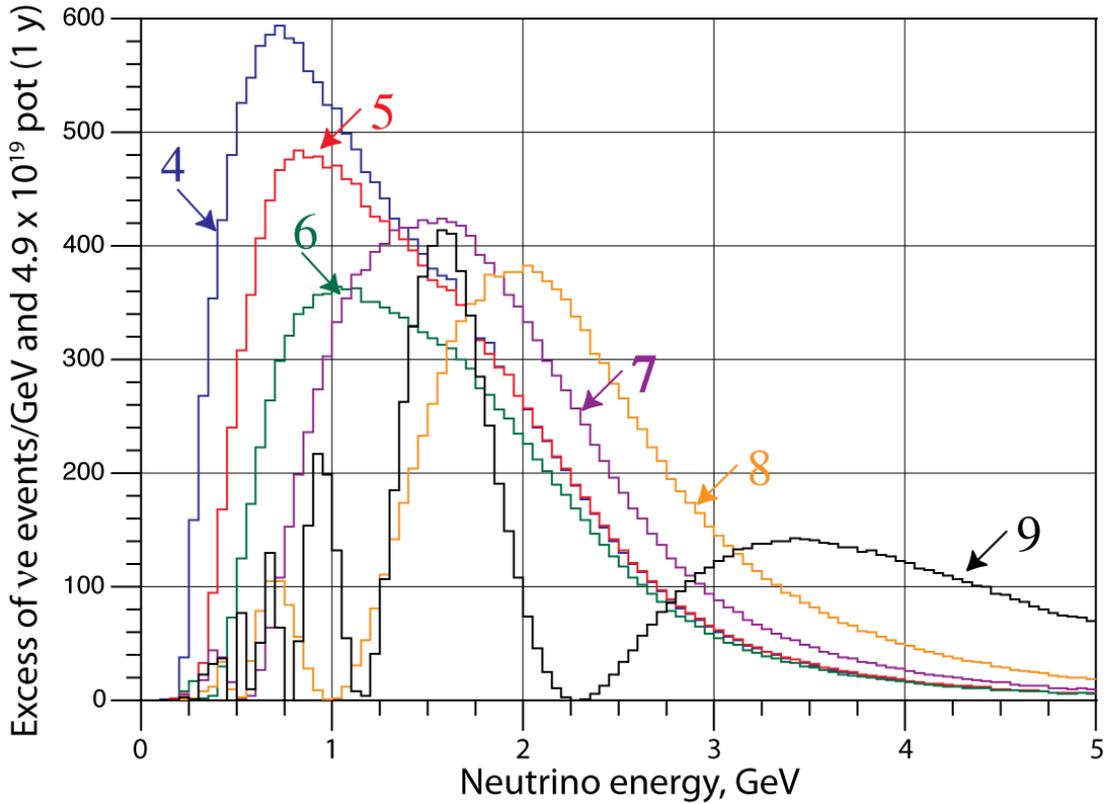

**Figure 3.** Expected excess of events for a sterile $\nu_\mu \to \nu_e$ oscillation signal at a distance d=1600 m (far position of the ICARUS detector) for $4.5 \times 10^{19}$ p.o.t. (1 y) and several different values of ($\Delta m^2$, $\sin^2(2\theta)$). Labels 4 ÷ 9 refer to the values in the Table.

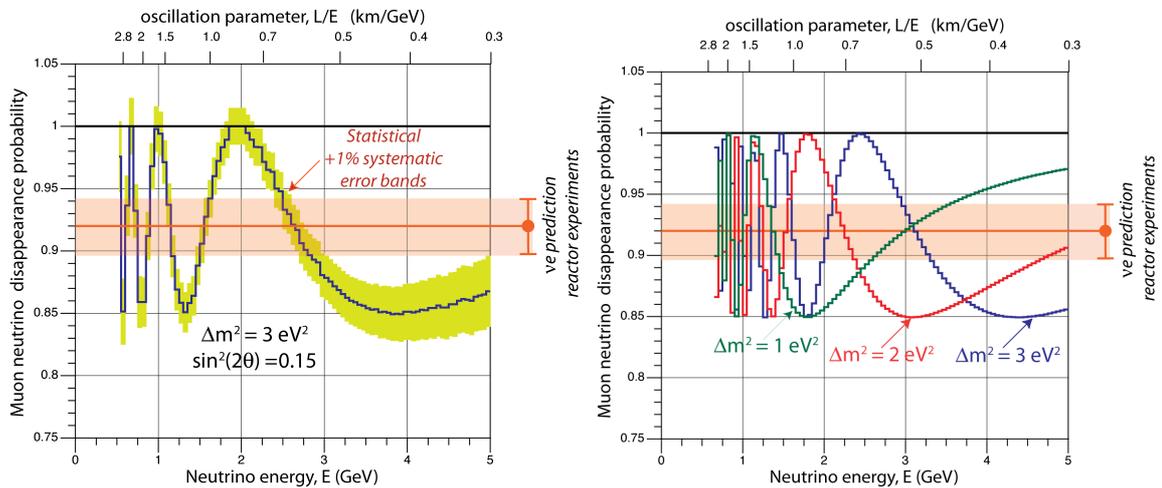

**Figure 4a and Figure 4b.** Muon oscillation disappearance in the LAr detector at d = 1.6 km, E < 5 GeV and $4.5 \times 10^{19}$ p.o.t. (1 y). The disappearance asymptotic probability for very large L/E has been chosen in accordance with the one observed from reactor experiments and the $\bar{\nu}_e$ channel [9], R = 0.927± 0.023. Graph 4a on the left show statistical and 1% systematic errors for $\Delta m^2$= 3 eV$^2$ and $\sin^2(2\theta)$ = 0.15) for d=1600 m and $4.5 \times 10^{19}$ p.o.t. (1 y).